\documentclass{article}  
\usepackage{graphicx, subfig, caption}
\usepackage{natbib}

\title{Leo5 Is a Z Cam Type Dwarf Nova}
\author{Patrick Wils$^1$, Tom Krajci$^2$, Mike Simonsen$^3$\\
\small
 $^1$Vereniging Voor Sterrenkunde, Belgium;  
email: patrickwils@yahoo.com \\
\small
 $^2$Astrokolkhoz, New Mexico, USA;  
email: tom\_krajci@tularosa.net \\
\small
 $^3$American Association of Variable Star Observers, USA;  
email: mikesimonsen@aavso.org \\
      }
\date{March 2011}

\begin{document}
\maketitle 

\begin{abstract}
Photometry of Leo5 = 1H 1025+220 show that it is a dwarf nova of the Z Cam subtype.  Two long standstills have been observed in the last five years.
\end{abstract}

\section{Introduction}

The Z Cam subtype is a rare class of dwarf novae, characterized by so-called standstills at an intermediate brightness level
between the outburst maximum and the quiescence state.  
These standstills may last for a number of weeks to years, when the object does not change in brightness very much.
The orbital periods of Z Cam type dwarf novae have been found to be always above the period gap for cataclysmic variables.
Less than 40 genuine members of the class are known \citep{ZCamPaign}.


Leo5 = 1H\,1025+220 = SDSS\,J102800.07+214813.5 has been a little studied cataclysmic variable (CV), 
discovered by Remillard et al. in the course of the HEAO-1 x-ray survey \citep[see ][]{Downes}.
It was classified as a nova-like variable and  
confirmed to be a CV spectroscopically by \citep{Munari}, and more recently also by the Sloan Digital Sky Survey \citep[SDSS; ][]{Szkody}.
\citet{Taylor} found it to be eclipsing with an orbital period of 3.506 hours.

\section{Observations}

\begin{figure*}
\centering
\includegraphics[width=13cm]{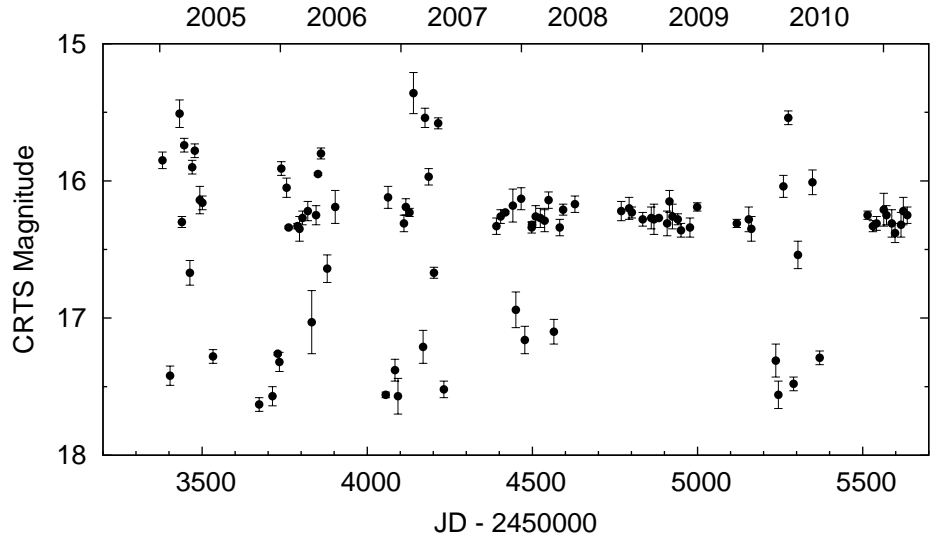}
\caption{Light curve of Leo5 from the Catalina Real-time Transient Survey.}
\label{LC1}
\end{figure*}

As part of its service to the study of transient objects, the Catalina Real-time Transient Survey \citep[CRTS; ][]{CRTS}
publicly provide their data for a number of CVs.  
For Leo5 data are available from 2005 to the present (Fig.~\ref{LC1}).
These data show a light curve which is typical of a low amplitude (magnitude 15.2-17.7), frequently erupting dwarf nova.  
On one occasion however, starting early 2008, outbursts seem to have ceased for more than a year, until the end of 2009, with Leo5 all the time at around magnitude 16.2.

\begin{figure*}
\centering
\includegraphics[width=13cm]{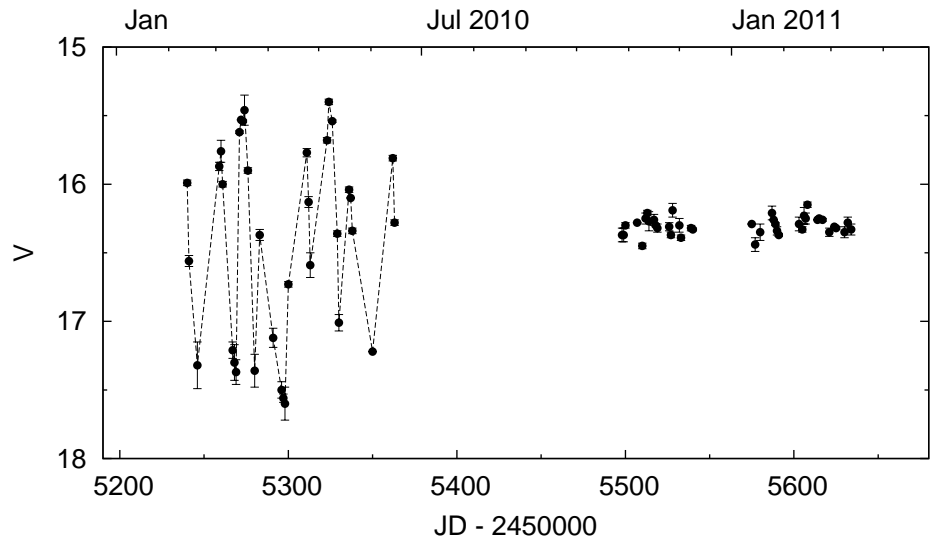}
\caption{$V$ light curve of Leo5 obtained through AAVSONet.  The data points from early 2010 have been connected with lines to lead the eye.}
\label{LC2}
\end{figure*}

CRTS data are fairly sparse in general (at best one night of four data points each week), so that more frequent observations were called for.
These were obtained in $B$ and $V$ through AAVSONet starting early 2010.
These data, presented in Fig.~\ref{LC2}, show the object again with frequent outbursts, varying between magnitude 15.5 and 17.5$V$, 
and spending little time in quiescence.
Although there are not enough observations to determine the length of an outburst cycle precisely, it is estimated to be about 20 days.
At the start of the 2010-2011 observing season however, the outbursts had ceased again, and Leo5 was at an intermediate magnitude of 16.2$V$.
This standstill has lasted until the present.
In hindsight the classification as a nova-like variable may have been partly due to the fact Leo5 is in standstill very often or for extended periods.

\section{Conclusion}

Leo5 = 1H\,1025+220 has been found to be a member of the rare class of Z Cam type dwarf novae, showing frequent and long standstills.
Because it is also an eclipsing system, Leo5 may provide more clues to the mechanism causing the standstills.
\\

\noindent{\bf Acknowledgements}\\
This study made use of NASA's Astrophysics Data System, and the SIMBAD and VizieR 
databases operated at the Centre de Donn\'ees astronomiques in Strasbourg (France).
The Catalina Real-time Transient Survey is acknowledged for making their data publicly available.
The data have been obtained through AAVSONet, the network of robotic telecopes of the AAVSO.


\begin{thebibliography}{6}

\bibitem[\protect\citeauthoryear{Downes \& Shara}{1993}]{Downes} {Downes}, R.A., Shara, M.M., 1993, PASP 105, 127

\bibitem[\protect\citeauthoryear{Drake et al.}{2009}]{CRTS} {Drake}, A.J., Djorgovski, S.G., Mahabal, A., Beshore, E., Larson, S., Graham, M.J., Williams, R., Christensen, E., Catelan, M., Boattini, A., Gibbs, A., Hill, R., Kowalski, R., 2009, ApJ 696, 870

\bibitem[\protect\citeauthoryear{Munari et al.}{1997}]{Munari} {Munari}, U., Zwitter, T., Bragaglia, A., 1997, A\&AS, 122, 495

\bibitem[\protect\citeauthoryear{Simonsen}{2011}]{ZCamPaign} {Simonsen}, M., 2010, JAAVSO 39, eJAAVSO 138

\bibitem[\protect\citeauthoryear{Szkody et al.}{2009}]{Szkody} {Szkody}, P., Anderson, S.F., Hayden, M., et al., 2009, AJ 137, 4011

\bibitem[\protect\citeauthoryear{Taylor}{1999}]{Taylor} {Taylor}, C., 1999, SW Sextantis stars, superhumps, and other phenomena in cataclysmic variables, Ph.D. thesis, Dartmouth College


\end{thebibliography}
\end{document}